\documentclass[11pt]{article}
\usepackage{amssymb}

\newcommand{\ket}[1]{\left | #1 \right \rangle}
\newcommand{\bra}[1]{\left \langle #1 \right |}
\newtheorem{theorem}{Theorem}

\begin{document}
\begin{center}
{\LARGE\bf Embedding classical into quantum computation}\\
\bigskip
{\normalsize Richard Jozsa}\\
\bigskip
{\small\it Department of Computer Science, University of
Bristol,\\ Merchant Venturers Building, Bristol BS8 1UB U.K.}
\end{center}

\begin{abstract}
We describe a simple formalism for generating classes of quantum
circuits that are classically efficiently simulatable and show
that the efficient simulation of Clifford circuits
(Gottesman-Knill theorem) and of matchgate circuits (Valiant's
theorem) appear as two special cases. Viewing these simulatable
classes as subsets of the space of all quantum computations, we
may consider minimal extensions that suffice to regain full
quantum computational power, which provides an approach to
exploring the efficacy of quantum over classical computation.
\end{abstract}
\bigskip

\section{Introduction}\label{intro}
The characterisation of the possibilities and limitations of quantum
computational power is one of the most interesting issues in quantum
information science. All of the early and best known quantum
algorithms \cite{NC} that exhibit an exponential time speed-up over
any known classical algorithm for the task, utilize properties of
the quantum Fourier transform modulo $N$. One may then develop
generalisations of these insights, studying Fourier transforms over
further abelian and non-abelian groups and invent associated
computational tasks such as the hidden subgroup problem and various
kinds of hidden shift problems. Around the years of 1997 and 1998
Thomas Beth, with memorable characteristic exuberance, was one of
the earliest workers in the subject to recognise the potential
possibilities of the abstract formalism of Fourier transforms for
novel quantum algorithms, and take up this line of development which
has now become an important cornerstone in our understanding.

Despite this seminal development it is probably fair to say that
apart from the Fourier transform formalism, no other similarly
fruitful quantum algorithmic primitive for exponential speed-up has
been identified. This motivates a need for alternative approaches to
exploring the efficacy of quantum vs. classical algorithms. One
interesting such approach is the identification and study of classes
of quantum computations that are classically efficiently simulatable
i.e. processes which although quantum, do {\em not} offer
computational benefit. Indeed the relation of classical to quantum
computation that emerges is intriguingly rich and multi-faceted --
(sub-) classical computation can be embedded into quantum
computation in many inequivalent ways. Given any such class of
simulatable quantum computations we may ask: what kind of added
(minimal) ingredient suffices to restore full quantum computational
power? In a sense, any such ingredient may be viewed as an
``essence'' of quantum computational power, albeit {\em relative} to
a given substrate of simulatable processes. In this talk we will
outline a formalism for providing simulatable classes of quantum
circuits and discuss two examples -- the Gottesman-Knill theorem for
Clifford circuits and Valiant's theorem for simulation of matchgate
circuits. These examples will show that the added ingredient above
can be strikingly trivial, especially if thought of as a competitor
to the oft-quoted blanket attribution of quantum computational power
to the enigmatic phenomenon of entanglement.

\section{Classically simulatable quantum computations}
\label{clsimsect} We focus on comparing and contrasting two theorems
of classical simulation {\em viz.} the Gottesman-Knill theorem for
Clifford circuits \cite{NC,cliffgp} and Valiant's theorem
\cite{valclsim,jm08} for simulation of matchgate circuits. At first
sight these appear to be very different in their content and
provenance but we will outline a proof method that reveals a formal
similarity between the two results.

The Gottesman-Knill (GK) theorem arose out of the development of the
so-called stabiliser formalism for the theory of quantum error
correction \cite{NC}. Let $H$ denote the 1-qubit Hadamard gate, $P$
the 1-qubit phase gate $P= {\rm diag}(1,i)$ and $CZ$ the 2-qubit
controlled$-Z$ gate $CZ={\rm diag}(1,1,1,-1)$. These gates and
arbitrary circuits of them on $n$ qubits are called {\em Clifford}
operations on $n$ qubits. Our adopted version (slightly modified
from the original, c.f. also \cite{cjl}) of the GK theorem is the
following.

\begin{theorem}\label{gk} Consider any uniform (hence poly sized)
quantum circuit family comprising the gates $H,P$ and $CZ$ (i.e. a
Clifford circuit) such that:\\ (i) the input state is any product
state;\\ (ii) the output is a final $Z$ measurement on any single
qubit line.\\ Then the output may be classically efficiently
simulated.
\end{theorem}

More formally our notion of efficient classical simulation is the
following: given a description of the circuit on $n$ qubit lines,
the output probabilities may be classically computed to $k$ digits
in poly$(n,k)$ time.

Next we introduce the notion of ``matchgate'' and Valiant's
classical simulation theorem \cite{valclsim}, which arose
originally from considerations of counting perfect matchings in
graphs.

A matchgate \cite{valclsim,jm08} is defined to be any 2-qubit gate
$G(A,B)$ of the form (in the computational basis):
\begin{equation}\label{gab} G(A,B) = \left(
\begin{array}{cccc} p&0&0&q \\ 0&w&x&0 \\ 0&y&z&0 \\ r&0&0&s
\end{array} \right) \hspace{1cm} A = \left( \begin{array}{cc}
p&q \\ r&s \end{array} \right) \hspace{5mm} B= \left(
\begin{array}{cc} w&x \\ y&z \end{array} \right) \end{equation}
where $A$ and $B$ are both in $SU(2)$ or both in $U(2)$ with the
{\em same determinant}. Thus the action of $G(A,B)$  amounts to $A$
acting in the even parity subspace (spanned by $\ket{00}$ and
$\ket{11}$) and $B$ acting in the odd parity subspace (spanned by
$\ket{01}$ and $\ket{10}$).

Our version of Valiant's theorem (again slightly different from the
original version) is the following.
\begin{theorem}\label{valiant} Consider any uniform (hence poly-sized)
quantum circuit family comprising only $G(A,B)$ gates such that:\\
(i)
the $G(A,B)$ gates act on nearest neighbour (n.n.) lines only;\\
(ii) the input state is any product state;\\ (iii) the output is a
final measurement in the computational basis on any single line.\\
Then the output may be classically efficiently simulated.
\end{theorem}

Let us now return to the GK theorem and its proof ingredients. The
essential property of the class of gates used, i.e. Clifford gates,
is the following \cite{cliffgp}: if $C$ is any $n$-qubit Clifford
operation and $P_1\otimes \ldots \otimes P_n$ is any product of
Pauli matrices (i.e. $P_i=I,X,Y$ or $Z$ for each $i$) then the
conjugate $C^\dagger (P_1\otimes \ldots \otimes P_n)C = P_1'\otimes
\ldots \otimes P_n'$ is again a product of Pauli operations. Stated
more formally, if ${\cal P}_n$ is the group generated by all such
Pauli products on $n$ qubits then the n-qubit Clifford group is the
normaliser of ${\cal P}_n$ in the unitary group $U(2^n)$.

A standard proof (c.f. \cite{NC}) of the GK theorem (with a
computational basis input) proceeds by updating the stabiliser
description of the state through the course of the computation and
we get a description of the final state in addition to the output
probabilities. We adopt here a different approach \cite{cjl}.
Suppose (wlog) that the final measurement is on the first line,
having outputs 0,1 with probabilities $p_0, p_1$ respectively. Then
the difference $p_0-p_1$ is given by the expectation value of
$Z_1=Z\otimes I\otimes \ldots \otimes I$ in the final state
$C\ket{\psi_0}$:
\begin{equation}\label{diffs} p_0-p_1= \bra{\psi_0} C^\dagger Z_1
C\ket{\psi_0} \end{equation} This computation suffices to simulate
the output (as also $p_0+p_1=1$). Now $Z_1$ is clearly a product of
Pauli operations so $C^\dagger Z_1C$ also has the product form
$P_1\otimes \ldots \otimes P_n$ for Pauli operations $P_i$ (whose
identity can be determined in linear time by an update rule for
successive conjugations by the elementary gates in the circuit).
Hence if $\ket{\psi_0}=\ket{a_1}\ldots \ket{a_n}$ is any product
state we get
\begin{equation}\label{prods} p_0-p_1= \prod_{k=1}^n
\bra{a_k}P_k\ket{a_k} \end{equation} which can clearly be
calculated in time $O(n)$ (as a product of $n$ terms of fixed
size) giving an efficient (linear time) simulation of the Clifford
circuit.

The essential ingredients of the above proof are the following.\\
{\bf (SIM1)}: we have a set ${\cal S}_n$ of $n$-qubit operations
such that $\bra{\psi_0} S \ket{\psi_0}$ can be computed in poly$(n)$
time for any $S\in {\cal S}_n$ and any allowed input state
$\ket{\psi_0}$;\\ (For the GK theorem ${\cal S}_n$ is the
$n$-qubit Pauli group ${\cal P}_n$.)\\
{\bf (SIM2)}: we have a class ${\cal K}_n$ of unitary operations
such that $K^\dagger S K\in {\cal S}_n$ for all $S \in {\cal S}_n$
and $K\in {\cal K}_n$.\\ (For the GK theorem ${\cal K}_n$ is the
Clifford group ${\cal C}_n$.)

Then if $Z_1$ is in ${\cal S}_n$ for all $n$ (or can be expressed
in suitably simple terms using elements of ${\cal S}_n$, c.f.
later) then it follows (just as in the above outlined proof) that
circuits of gates from ${\cal K}_n$, with input state
$\ket{\psi_0}$ and output measurement of $Z$ on the first line,
can be classically efficiently simulated.

Note that this simulation result, resting on (SIM1) and (SIM2) does
not actually require any special group (or other algebraic)
structure on ${\cal S}_n$ or ${\cal K}_n$. For example, the fact
that ${\cal P}_n$ is a {\em group} is not needed at all in our proof
of the GK theorem in contrast to the usual proof resting on the
stabiliser formalism, depending heavily on the subgroup structure of
${\cal P}_n$.

Turning now to matchgates we will show that Valiant's theorem can
be understood as just another example of the above formalism with
a suitably clever choice of ${\cal S}_n$ and ${\cal K}_n$. For $n$
qubits we introduce the $2n$ Pauli product operators (omitting
tensor product symbols $\otimes$ throughout):
\begin{equation}\label{jwrep} \begin{array}{c}
c_1=X\,I\ldots I \hspace{3mm} c_3= Z\,X\,I\ldots I \hspace{2mm}
\cdots \hspace{2mm} c_{2k-1}= Z\ldots Z\,X\,I\ldots I \hspace{2mm}
\\  c_2=Y\,I\ldots I \hspace{3.5mm} c_4= Z\,Y\,I\ldots I \hspace{2mm}
\cdots \hspace{3mm}  c_{2k}\,\,\,\, = Z\ldots Z\,Y\,I\ldots I
\hspace{2mm}
\end{array}
\end{equation} where $X$ and $Y$ are in the $k^{\rm th}$ slot for
$c_{2k-1}$ and $c_{2k}$, and $k$ ranges from 1 to $n$. For ${\cal
S}_n$ we take the {\em linear span} of $c_1, \ldots , c_{2n}$
which is a $2n$-dimensional vector space (in contrast to the group
${\cal P}_n$). Since each $c_j$ is a product operator and a
general vector $v\in {\cal S}_n$ is a linear combination of only
$2n$ of them, it is clear that $\bra{\psi_0} v\ket{\psi_0}$ is
poly$(n)$-time computable if $\ket{\psi_0}$ is a product state
i.e. (SIM1) is satisfied.

Next we can verify by straightforward direct calculation that if $U$
is any n.n. $G(A,B)$ gate then $U^\dagger c_j U \in {\cal S}_n$ for
all $j$ so $U^\dagger v U \in {\cal S}_n$ for any $v\in {\cal S}_n$
i.e. property (SIM2) is satisfied. More explicitly note that if $U$
is a n.n. $G(A,B)$ gate, it applies to two {\em consecutive} qubit
lines so (from eq. (\ref{jwrep})) the part of $c_j$ that it ``sees''
can only be one of
\begin{equation} \label{alphas} \alpha_1=ZZ\hspace{3mm}
\alpha_2=ZX\hspace{3mm}\alpha_3=ZY\hspace{3mm}\alpha_4=XI\hspace{3mm}
\alpha_5=YI\hspace{3mm}{\rm or}\hspace{3mm}\alpha_6=II.
\end{equation} Then a straightforward calculation with 4 by 4
matrices shows that for each $i$, $G(A,B)^\dagger\alpha_i G(A,B)$
always returns a linear combination of allowable $\alpha_i$'s and
property (SIM2) follows immediately.

It is instructive to note that if we attempt to apply a $G(A,B)$
gate on {\em not} nearest-neighbour qubit lines then in addition to
the six terms in eq. (\ref{alphas}) we can get a {\em further}
possibility, namely $\alpha_7=ZI$ on the chosen two lines. But now
we can check that $G(A,B)^\dagger \alpha_7 G(A,B)$ does {\em not}
generally lie in the span of the allowed Pauli products at those
lines, and property (SIM2) is violated. This give a way of
understanding the curious n.n. requirement for $G(A,B)$ actions in
theorem \ref{valiant}, which has no analogue in the GK theorem (as
${\cal P}_n$ is defined by a uniformly local product requirement).

With properties (SIM1) and (SIM2) we can say that if $M$ is the
total operation of any n.n. matchgate circuit on $n$ lines then
$\bra{\psi_0} M^\dagger DM\ket{\psi_0}$ is poly$(n)$-time computable
for any $D\in {\cal S}_n$. To complete our simulation theorem we
want to set $D=Z_k = I \ldots I\ Z  I  \ldots I$ (i.e. $Z$ on the
$k^{\rm th}$ line) to obtain $p_0-p_1$ for a measurement on the
$k^{\rm th}$ line.  In the GK theorem with ${\cal S}_n = {\cal P}_n$
we had $Z_k \in {\cal P}_n$ directly. In the present case we do not
have $Z_k\in {\cal S}_n$ but looking at eq. (\ref{jwrep}) we see
that $Z_1=-ic_1c_2$ and generally $Z_k=-ic_{2k-1}c_{2k}$. Then, for
example,
\begin{equation} M^\dagger Z_1 M = -i M^\dagger c_1c_2 M =
-i (M^\dagger c_1 M)( M^\dagger c_2 M) \end{equation} and each
bracket in the last expression is a linear combination of $c_j$'s.
Thus $p_0-p_1=\bra{\psi_0}M^\dagger Z_1 M \ket{\psi_0}$ has the
form $-i \sum_{ij} a_ib_j \bra{\psi_0}c_ic_j \ket{\psi_0}$. Since
the $c_i$'s are product operators, so are the $O(n^2)$ product
terms $c_ic_j$ in the final sum. Hence $p_0-p_1$ is again
poly$(n)$-time computable but now we have $O(n^2)$ terms instead
of the previous $O(n)$ terms in the sum. This completes a proof
outline of Valiant's theorem \ref{valiant}.

\section{Extensions of simulatable circuits}\label{extensions}
We may now view Clifford circuits and matchgate circuits as two
``islands'' of quantum processes in the space of all quantum
computations, that offer no computational time benefit over
classical computations. As such, it is interesting to try to
characterise their relationship to the whole and one approach is to
consider what (minimal) extra ingredient suffices to expand their
computational power to regain full universal efficient quantum
computation.

In the case of Clifford circuits it is well known (e.g. see
\cite{NC}) that the inclusion of the phase gate
$\sqrt{P}=\rm{diag}(1, e^{i\pi/4})$ suffices, and more generally,
(using a result of Shi \cite{shi}, noting that $CNOT$ is a
Clifford operation), the inclusion of essentially any single extra
non-trivial 1-qubit gate will suffice.

For the case of matchgate circuits we have the following
intriguing result.
\begin{theorem}\label{two} Let $C_n$ be any uniform family of
quantum circuits with output given by a $Z$ basis measurement on
the first line. Then $C_n$ may be simulated by a circuit of
$G(A,B)$ gates acting on n.n. or {\em next} n.n. lines only (i.e.
on line pairs at most distance 2 apart) with at most a constant
factor increase in the size of the circuit.
\end{theorem}
A proof of this theorem may be found in \cite{jm08} and here we just
make a few remarks. Comparing theorems \ref{valiant} and \ref{two}
we see that the gap between classical and full quantum computational
power can be bridged by a very modest use of a seemingly innocuous
resource viz. the ability of matchgates to act on next n.n. --
instead of just n.n. -- qubit lines. Equivalently this may be
characterised by use of the $SWAP$ operation (on n.n. lines) in a
very constrained context where {\em ladders} of consecutive $SWAP$s
(which would allow 2-qubit gates to act on arbitrarily distant
lines) are not even allowed. From this perspective, the power of
quantum (over classical) computation is attributable to the mere
inclusion of such isolated single $SWAP$ gates. The result becomes
perhaps even more striking if we note that $SWAP$ itself is very
close to being expressible in the allowed $G(A,B)$ form. Indeed
$SWAP = G(I,X)$ and fails only through a mere minus sign in ${\rm
det}\,X = - {\rm det}\,I$. Thus if we drop the ${\rm det}A= {\rm
det}B$ condition in eq. (\ref{gab}), then the resulting $G(A,B)$
gates acting on n.n. lines become efficiently universal for quantum
computation.

Is it conceivable that the passage from n.n. to next-n.n. use of
$G(A,B)$ gates may be achieved while maintaining classical
simulatability? We may argue on formal complexity theoretic
grounds that this is highly implausible. Indeed it is shown in
\cite{jm08} that the classical complexity classes NP and PP (cf.
\cite{papadim}) would then become classically poly-time decideable
i.e. we would get P$=$NP$=$PP (as well as P$=$BQP). Thus an extra
supra-classical computational power {\em must} be associated to
the single distance extension of the range of n.n. 2-qubit
$G(A,B)$ gates in general matchgate circuits,  if these classical
computational complexity classes are to be unequal.

\section{Concluding remarks}\label{conclude}
>From the viewpoint of (SIM1) and (SIM2) we see a formal similarity
between the GK theorem and Valiant's theorem although these
results arose historically from very different considerations.
This suggests that we might be able to construct further
interesting classes of classically simulatable circuits by simply
taking other choices of ${\cal S}_n$ and identifying a suitable
associated ${\cal K}_n$. However ``interesting'' pairs $({\cal
S}_n,{\cal K}_n)$ appear to be difficult to invent -- the known
examples arising as outcomes of some prior elaborate underlying
mathematical structures. In the GK case we have the identification
of the Clifford group via a lengthy argument with group theoretic
ingredients (see e.g. appendix in \cite{clark}) applied to the
Pauli group ${\cal P}_n$ which is a well known structure in the
subject.

However in the case of Valiant's theorem, how might we initially
come upon this result, and guess the choice for ${\cal S}_n$ that we
used (i.e. eq. (\ref{jwrep}) and its linear span)!? Actually the
operators in eq. (\ref{jwrep}) are well known in physics -- they
comprise the so-called Jordan-Wigner representation \cite{jwigrep}
that appears in the theory of non-interacting fermions. The
connection between Valiant's theorem and simulation of free fermions
was recognised by Knill \cite{kn} and Terhal and DiVincenzo
\cite{terdiv} and our proof of Valiant's theorem above is a
re-writing of this connection. A more formal mathematical treatment
(albeit without reference to fermions) based on abstract properties
of the mathematical structure of Clifford algebras is given in
\cite{jm08} which also clarifies the appearance of matchgates as
normalisers of the linear part of the Clifford algebra, leading to
property (SIM2). We will not elaborate here on these further
ingredients (detailed in \cite{jm08}) except to point out that again
here, we have a significant underlying theory leading to the choice
of ${\cal S}_n$ and the identification of its associated normalisers
${\cal K}_n$. Perhaps an intuitive signal feature of such an
underlying theory is some construction that could potentially
produce an exponentially large structure but surprisingly remains
only polynomially complex. In the case of the Pauli group ${\cal
P}_n$, conjugation by arbitrary $V\in U(2^n)$ can generate general
$n$-qubit matrices for which the calculation of the expectation
value in eq. (\ref{diffs}) becomes exponentially inefficient. But
the special case of $V$ being Clifford guarantees a polynomial
simplicity via the preserved product structure. In the case of the
$c_i$'s of eq. (\ref{jwrep}), conjugation by an arbitrary $V\in
U(2^n)$ leads to a general element of the full Clifford algebra
generated by the the $c_i$'s \cite{jm08} -- a space of exponential
dimension $2^{2n}$ -- but again the special case of n.n. matchgates
(associated to a theory of quadratic hamiltonians \cite{jm08})
guarantees that the conjugates remain in the polynomially small
subspace of linear elements of the full Clifford algebra. It is an
interesting open problem to exhibit further examples of such
simplifications and of our formalism (SIM1), (SIM2), that may
already exist within the literature of the theory of some yet more
general kind of algebraic structure.

\noindent {\bf Acknowledgements.} This work was supported in parts
by the EC networks QICS and QAP and by EPSRC QIP-IRC.  The author
also acknowledges Akimasa Miyake for the collaborative work
\cite{jm08} which is related closely to the discussion in this
paper.

\end{document}